# Mechanical properties of freely suspended semiconducting graphene-like layers based on $MoS_2$

**Andres Castellanos-Gomez**[1,2,\*], Menno Poot[1,+], Gary A. Steele[1], Herre S.J. van der Zant[1], Nicolás Agraït[2,3] and Gabino Rubio-Bollinger[2]

[1] Kavli Institute of Nanoscience, Delft University of Technology, Lorentzweg 1, 2628 CJ Delft, The Netherlands.

[2] Departamento de Física de la Materia Condensada (C–III). Universidad Autónoma de Madrid, Campus de Cantoblanco, E-28049 Madrid, Spain.

[3] Instituto Madrileño de Estudios Avanzados en Nanociencia IMDEA-Nanociencia, E-28049 Madrid, Spain.

[+] Present address: Yale University. Department of Engineering Science. Becton 215, 15 Prospect St. New Haven, CT 06520. USA.

a.castellanosgomez@tudelft.nl

**ABSTRACT**

We fabricate freely suspended nanosheets of $MoS_2$ which are characterized by quantitative optical microscopy and high resolution friction force microscopy. We study the elastic deformation of freely suspended nanosheets of $MoS_2$ using an atomic force microscope. The Young's modulus and the initial pre-tension of the nanosheets are determined by performing a nanoscopic version of a bending test experiment. $MoS_2$ sheets show high elasticity and an extremely high Young's modulus (0.30 TPa, 50% larger than steel). These results make them a potential alternative to graphene in applications requiring flexible semiconductor materials.

**BACKGROUND**

The application of graphene in semiconducting devices is hindered by its lack of a bandgap. Up to now, two different strategies have been employed to fabricate semiconducting two-dimensional crystals: opening a bandgap in graphene [1-3] or using another two-dimensional crystal with a large intrinsic bandgap [4]. Atomically thin molybdenum disulphide ($MoS_2$), a semiconducting transition metal dichalcogenide, have recently attracted a lot of attention due to its large intrinsic bandgap of 1.8 eV and high mobility $\mu > 200$ $cm^2V^{-1}s^{-1}$ [5, 6]. In fact, $MoS_2$ has been employed to fabricate field-effect-transistors with high on/off ratios [5], chemical sensors [7] and logic gates among other things [8]. Nevertheless, the study of the mechanical properties of this nanomaterial (which will dictate its applicability in flexible electronic applications) have just begun [9, 10]. In a previous work, we studied the mechanical properties of freely suspended MoS$_2$ nanosheets using a bending test experiment performed with the tip of an atomic force microscope (AFM) [10].

Here we perform a more detailed characterization of the fabricated nanosheets by quantitative optical microscopy and high resolution friction force microscopy and we extend the study of the mechanical properties to a larger number of MoS$_2$ nanosheets (with thicknesses in the range of 5 to 25 layers) to improve the robustness of our statistical analysis. We present force *vs.* deformation curves measured not only by pushing the nanosheets (as usual) but also by pulling them, demonstrating that for moderate deformations pushing and pulling the nanosheets is equivalent. These measurements allow for the simultaneous determination of the Young's modulus (*E*) and the initial pre-tension (*T*) of these MoS$_2$ nanosheets.

**MATERIALS AND METHODS:**

Although atomically thin MoS$_2$ crystals can be fabricated by scotch-tape-based micromechanical cleavage [11], this procedure can leave traces of adhesive. Thus, it is preferable to use an all-dry technique based on poly (dimethyl)-siloxane (PDMS) stamps which has been successfully employed to fabricate ultra-clean atomically thin crystals of graphene [12], graphene nanoribbons [13], NbSe$_2$, MoS$_2$ [14] and muscovite mica [15]. In order to fabricate freely suspended atomically thin MoS$_2$ flakes the cleaved flakes are transferred to a pre-patterned oxidized silicon wafer [16] with circular holes 1.1 µm in diameter and 200 nm deep.

After fabrication, an optical microscope (*Nikon Eclipse LV100*) is used to identify MoS$_2$ flakes at first glance. In fact, ultrathin MoS$_2$ flakes deposited onto a silicon wafer with a 285 nm thick SiO$_2$ capping layer can be easily identified by optical microscopy. Figure 1(a) shows a chart of the expected color of MoS$_2$ flakes with different thicknesses when they are laying on the surface or covering a hole. The expected color has been calculated with a Fresnel law-based model, employing the refractive index of MoS2 in Ref. [14] and the response of the camera as indicated in Ref.[17]. The topography of selected flakes is then studied by contact mode atomic force microscopy (AFM) to avoid possible artefacts in the flake thickness measurements [18]. Figure 1(b) and (c) are respectively an optical micrograph and a contact mode AFM topography of an 8 layers thick MoS$_2$ flake deposited onto a 285 nm SiO$_2$/Si pre-patterned substrate.

Additionally, high resolution contact mode AFM measurements can provide lattice resolution even in the suspended region of the MoS$_2$ flakes which demonstrates the very clean nature of our fabrication technique. Figure 1(d-e) shows two lateral force maps (friction images) obtained in the suspended and the supported part of the MoS$_2$ flake shown in Figure 1(c). The atomic resolution can be better resolved in the suspended region of the MoS$_2$ flake (Figure 1d) while in the supported part the frictional force image mainly follows parallel stripes (Figure 1e). We have employed a two-dimensional Tomlinson model [19] to simulate the frictional force image measured in the supported part of the nanolayer (see inset in Figure 1e), finding a remarkable qualitative agreement. Interestingly, by reducing a 25% in the depth of the surface potential employed in the simulation the calculated friction force image qualitatively matches the one measured in the suspended part of the MoS$_2$ nanomembrane (Figure 1d). This difference in the frictional force image can be due to a slight modification of the MoS$_2$ lattice induced by the pre-tension of the suspended part of the sheet. However, a detailed analysis of the tension dependence of frictional force images and their interpretation, although interesting, is beyond the scope of this work.

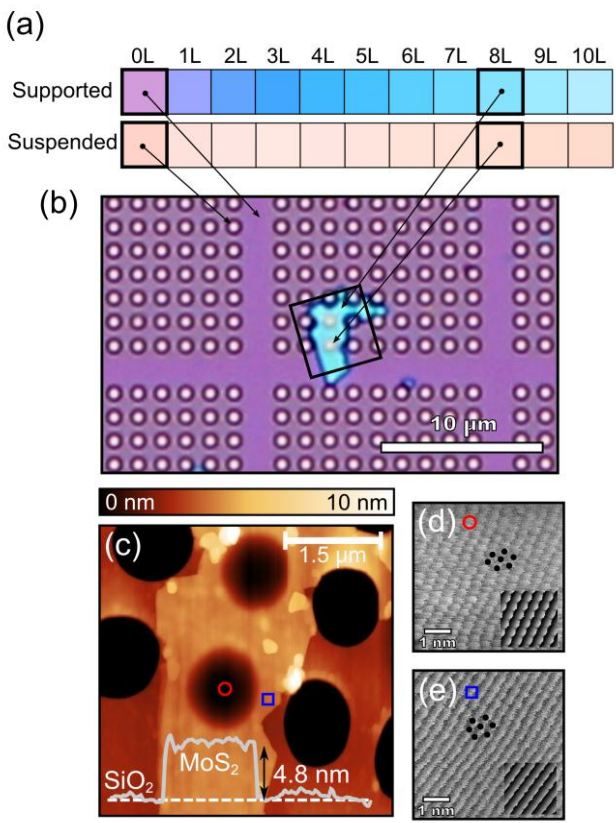

**Figure 1. Identification and characterization of freely suspended MoS$_2$ nanosheets.** (a) Color chart displaying the calculated color for MoS$_2$ nanosheets with different number of layers laying on the substrate (supported) or covering a hole (suspended). (b) Optical micrograph of a 4.8 nm thick (8 layers) MoS$_2$ flake deposited on top of a 285 nm SiO$_2$/Si substrate pre-patterned with an array of holes 1.1 μm in diameter. Even though the flake covers two holes, it is thin (and transparent) enough to permit optical identification of the covered holes, which present a slightly different color from the uncovered holes, as predicted by the color chart shown in (a). (c) Contact mode AFM topography of the region marked by the rectangle in (b). (Inset) Topographic line profile acquired along the dashed line in (c). (d) and (e) Raw friction forward images acquired in contact mode AFM in a suspended and a supported region, marked by a red circle and a blue square in (c) respectively. The insets in (d) and (e) show two friction images simulated with a two-dimensional Tomlinson model. Both images have been simulated employing the same crystal lattice and orientation but different depth of the potential well (see text).

**RESULTS AND DISCUSSION**

Once the suspended nanosheet under study is identified and characterized, we measure its elastic mechanical properties by using the AFM tip to apply a load cycle in the center of the suspended region of the nanosheet while its deformation is measured, as shown in the inset Figure 2(a) When the tip and sample are in contact, the elastic deformation of the nanosheet ($\delta$), the deflection of the AFM cantilever ($\Delta z_c$) and the displacement of the scanning piezotube of the AFM ($\Delta z_{piezo}$) are related by

$$\delta = \Delta z_{piezo} - \Delta z_c \qquad (1)$$

The force applied is related to the cantilever deflection as $F = k_c \cdot \Delta z_c$, where $k_c$ is the spring constant of the cantilever ($k_c = 0.75 \pm 0.20$ N/m [20]).

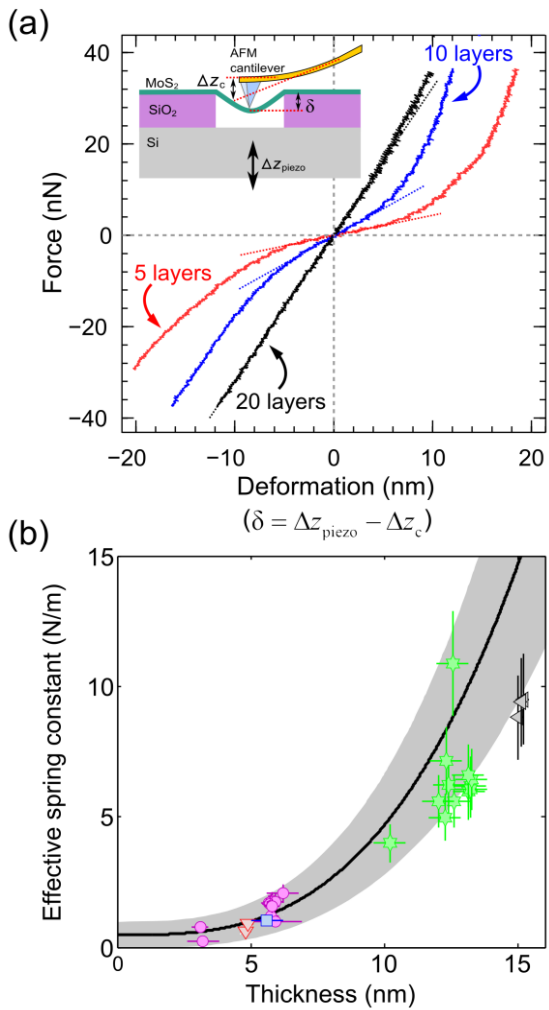

Figure 2. **Bending test experiment on suspended ultrathin MoS$_2$ sheets.** (a) Force *vs.* deformation traces measured by pushing and pulling at the center of the suspended part of MoS$_2$ nanosheets with 5, 10 and 20 layers in thickness. The slope of the traces around zero deformation is marked by a dotted line. (b) Effective spring constant as a function of the thickness measured for 31 MoS$_2$ suspended nanosheets with thickness ranging from 25 down to 5 layers. Data points sharing color and symbol correspond to suspended nanosheets from the same MoS$_2$ flake. The solid blak line shows the calculated relationship with expression (3) using $E = 0.30$ TPa and $T = 0.15$ N/m. The gray area around the solid black line indicates the uncertainty of $E$ and $T$: $\Delta E = \pm 0.10$ TPa and $\Delta T = \pm 0.15$ N/m.

Figure 2(a) shows 3 different deformation *versus* force ($F(\delta)$ hereafter), measured for nanosheets with 5, 10 and 20 layers in thickness, not only by pushing the sheets but also by pulling them. For small deformations, these $F(\delta)$ traces are linear with a slope that defines the effective spring constant of the nanolayer ($k_{eff}$) [21]

$$k_{\text{eff}} = \left.\frac{\partial F}{\partial \delta}\right|_{\delta=0} = \frac{4\pi E}{3(1-v^2)} \cdot \left(\frac{t^3}{R^2}\right) + \pi T \qquad (2)$$

with $v$ the Poisson's ratio ($v = 0.125$, ref. [22]), $t$ the thickness and $R$ the radius of the nanosheet. As the effective spring constant depends on both the Young's modulus and the pre-tension constant one cannot separately determine these values just from the slope of a $F(\delta)$ trace. To independently determine $E$ and $T$, however, one can use the thickness dependence of the effective spring constant. Indeed, according to expression (2), the first term (which accounts for the bending rigidity of the layer) strongly depends on the sheet thickness while the second one (which accounts for the initial pre-tension) is thickness independent. Fitting the measured $k_{\text{eff}}$ *versus* thickness to expression (2) one can determine $E$ and $T$. Figure 2(b) shows the measured $k_{\text{eff}}$ as a function of the thickness of 31 different MoS$_2$ layers and the fit to the experimental data using: $E = 0.30 \pm 0.10$ TPa and $T = 0.15 \pm 0.15$ N/m.

This Young's modulus value is extremely high, only one third lower than exfoliated graphene (one of the stiffest materials on Earth with $E = 0.8 - 1.0$ TPa) [23, 24] and comparable to other 2D crystals such as graphene oxide (0.2 TPa) [25] or hexagonal boron nitride (0.25 TPa) [26]. It is also remarkable that the $E$ value is restrained between 0.2 TPa and 0.4 TPa, indicating a high homogeneity of the MoS$_2$ flakes, which is much smaller than the one observed for graphene (0.02 – 3 TPa) [27] or graphene oxide (0.08 – 0.7 TPa) [25]. The high Young's modulus of the ultrathin MoS$_2$ flakes ($E = 0.30 \pm 0.10$ TPa compared to the bulk value $E_{\text{bulk}} = 0.24$ TPa [28]) can be explained by a low presence of stacking faults. Indeed the thinner the nanosheet the lower the presence of stacking faults, allowing the study of the intrinsic mechanical properties of the material.

**CONCLUSION**

We have studied the mechanical properties of ultrathin freely suspended MoS$_2$ nanosheets 5 to 25 layers thick. The mean Young's modulus of these suspended nanosheets, $E = 0.30 \pm 0.07$ TPa, is extremely high, and they present low pre-strain and high strength, being able to stand elastic deformations of tens of nanometers elastically without breaking. In summary, the low pre-tension and high elasticity and Young's modulus of these crystals makes them attractive substitutes or alternatives for graphene in applications requiring flexible semiconductor materials.

**ABBREVIATIONS**

AFM: Atomic Force Microscope, PDMS: poly (dimethyl)-siloxane.

## ACKNOWLEDGMENT

This work was supported by MICINN (Spain) through the programs MAT2011-25046 and CONSOLIDER-INGENIO-2010 'Nanociencia Molecular' CSD-2007-00010, Comunidad de Madrid through program Nanobiomagnet S2009/MAT-1726 and the European Union (FP7) through the programs RODIN.

## AUTHORS' CONTRIBUTIONS

ACG carried out the transfer and characterization of $MoS_2$ nanolayers and the bending test measurements. MP fabricated the pre-patterned substrates. ACG and GRB participated in the design and coordination of the experiments and designed the manuscript layout. MP, GAS, HSJvdZ and NA participated in the drafting of the manuscript and helped with the interpretation of the data. All authors read and approved the final manuscript.

## COMPETING INTERESTS

The authors declare that they have no competing interests.